\documentclass{article}
\usepackage{spconf}
\usepackage{amsmath,amssymb,graphicx,hyperref}
\usepackage{enumitem}

\usepackage{hyperref}
\usepackage{svg}
\usepackage{caption}
\usepackage{subcaption}
\usepackage{booktabs}

\newcommand{\cat}[2]{\underset{#1} {\overset{#2} {\mathbin\Vert}}}

\title{Adapting Diarization-Conditioned Whisper for End-to-End Multi-Talker Speech Recognition\\
}
\makeatletter
\def\@name{ 
\emph{Martin Kocour$^{\star,\dagger}$ 
\qquad Martin Karafiat$^{\star}$ 
\qquad Alexander Polok$^{\star}$
}\\
\emph{Dominik Klement$^{\star}$ 
\qquad Lukáš Burget$^{\star}$}
\qquad Jan Černocký$^{\star}$}
\makeatother
\address{$^{\star}$ Speech@FIT, Brno University of Technology, Czechia \\
            $^{\dagger}$ Filevine, USA \\
}
\copyrightnotice{\copyright\ IEEE 2026}
\toappear{Submitted to {\it ICASSP2026,
    May 04-08, 2026, Barcelona, Spain}}
    
\begin{document}
\ninept
\maketitle
\begin{abstract}
\vspace{-.5em}
We propose a speaker-attributed (SA) Whisper-based model for multi-talker speech recognition that combines target-speaker mo\-de\-ling with serialized output training (SOT).
Our approach leverages a Diarization-Conditioned Whisper (DiCoW) encoder to extract target-speaker embeddings, which are concatenated into a single representation and passed to a shared decoder.
This enables the model to transcribe overlapping speech as a serialized output stream with speaker tags and timestamps.
In contrast to target-speaker ASR systems such as DiCoW, which decode each speaker separately, our approach performs joint decoding, allowing the decoder to condition on the context of all speakers simultaneously.
Experiments show that the model outperforms existing SOT-based approaches and surpasses DiCoW on multi-talker mixtures (e.g., LibriMix).
\end{abstract}

\vspace{-1em}
\section{Introduction}
\vspace{-1em}
Automatic speech recognition (ASR) has seen remarkable progress over the past decade, driven by large-scale datasets, powerful neural architectures, and self-supervised learning techniques~\cite{hsu2021hubert,chen2022wavlm}. Most ASR systems, however, have traditionally assumed single-speaker, clean speech conditions typical of voice search, dictation, and other laboratory-controlled benchmarks, where recent models have achieved near-human performance, even on long-form audio.

In contrast, real-world conversations are inherently multi-speaker, often with overlapping speech, dynamic turn-taking, and background noise. These factors complicate transcription, particularly in multi-party dialogues, where speaker turns frequently interleave and overlap. Consequently, the research focus has expanded toward multi-talker ASR, often incorporating speaker diarization~\cite{meng23b_interspeech, meng24c_interspeech, kanda21b_interspeech,han25_icassp,zmolikova20_chime} to segment and attribute speech to individual speakers. The series of CHiME challenges~\cite{cornell2024chime}, which involved distant-microphone recordings of conversations, underscored the limitations of conventional ASR in such settings, highlighting the need for systems that can robustly handle overlap and noise.

Recent advances in multi-talker ASR (MT-ASR) have focused on directly transcribing overlapping speech while handling speaker assignment. Permutation-invariant training~\cite{yu2017permutation} addresses the issue of speaker label ambiguity but suffers from factorial complexity, motivating successors such as HEAT~\cite{raj23_surt2,lu21_icassp} or Sortformer~\cite{park2025sortformer}, which approximates optimal assignments. Serialized output training (SOT)~\cite{kanda20b_interspeech} simplifies decoding with speaker change tokens but lacks explicit speaker modeling. To better leverage speaker cues, Diarization-Conditioned Whisper (DiCoW)~\cite{polok2024dicowdiarizationconditionedwhispertarget,polok2026sedicowselfenrolleddiarizationconditionedwhisper} adapts Whisper with lightweight diarization-conditioned modules, DNCASR~\cite{zheng-etal-2025-dncasr} and SLIDAR~\cite{cornell2024one} integrate diarization and ASR within a unified model, and Microsoft’s Whisper extension~\cite{li23o_interspeech} injects speaker change tokens to improve overlap handling.

In this work, we build on prior advances by proposing a modified Whisper-based architecture that unifies target-speaker ASR (TS-ASR) and serialized output training (SOT). Central to our approach is the integration of a Diarization-Conditioned Whisper (DiCoW) encoder, pretrained for TS-ASR, which uses diarization information to focus on individual speakers. For each speaker in the recording, the encoder produces a dedicated representation - referred to as a \textit{speaker-channel} - that captures speaker-specific acoustic features. These embeddings are then jointly fed into a single shared decoder, enabling simultaneous decoding across overlapping speakers. The decoder generates serialized transcriptions that include both speaker tags and timestamps. Unlike conventional TS-ASR systems that decode each speaker independently, our model performs joint decoding conditioned on the global conversational context. This holistic design enhances robustness in highly overlapped speech scenarios, as demonstrated through evaluations on both synthetic mixtures (e.g., Libri2Mix)~\cite{Cosentino2020LibriMixAO} and real-world multi-speaker recordings such as AMI~\cite{ami_corpus} and NOTSOFAR~\cite{abramovski2025summarynotsofar1challengehighlights}.
This work aims to advance multi-talker ASR by integrating diarization-based conditioning with serialized output training in a unified architecture. 



\vspace{-1em}
\section{Methods}
\vspace{-.5em}


\vspace{-0.25em}
\subsection{Speech recognition with Whisper}
\vspace{-0.5em}

We build on the Whisper model \cite{radford2023whisper}, which is a multilingual encoder-decoder ASR system trained on a large collection of weakly labeled data. Whisper follows the attention-based encoder-decoder (AED) architecture \cite{chorowski2015_aed,chan2015_las}, with both the encoder and decoder composed of Transformer blocks \cite{vaswani2017attention}.

The audio encoder takes as input log mel-filterbank features $\mathbf{X} \in \mathbb{R}^{d_f \times 2T}$ and maps them into hidden embeddings:
\begin{equation}
\mathbf{H} = \text{Encoder}(\mathbf{X}), \quad \mathbf{H} \in \mathbb{R}^{d_m \times T}
\end{equation}
where $d_f$ and $2T$ are the feature dimension and frame count of the input, and $d_m$ and $T$ are the corresponding dimensions of the encoder output, since the input is subsampled in initial convolutional layers by factor of $2$.

The decoder generates the output tokens autoregressively, conditioning on the previously predicted tokens $Y_{1:n-1}$, a task-specific prefix $\mathbf{t}$, and the encoder output $\mathbf{H}$:
\begin{equation}
\mathbf{o}_n = \text{Decoder}(\mathbf{t}, Y_{1:n-1}, \mathbf{H}),\quad \mathbf{o}_n \in \mathbb{R}^{|\mathcal{V} \cup \mathcal{W}|}
\end{equation}
where $\mathbf{o}_n$ is the output distribution over the vocabulary $\mathcal{V}$ and set of timestamps $\mathcal{W}$ at step $n$.

Although Whisper performs well on a wide range of single-talker domains, it is not explicitly trained for multi-talker ASR and lacks the ability to perform speaker attribution. Consequently, its performance deteriorates in overlapping speech scenarios~\cite{polok2024dicowdiarizationconditionedwhispertarget}. To address this, we build on DiCoW and fine-tuned Whisper with components designed for speaker-aware modeling and multi-talker output generation.

\vspace{-1em}
\subsection{Target-speaker conditioning in DiCoW}
\vspace{-0.5em}

In DiCoW, the encoder is adapted for target-speaker ASR using diarization masks. These masks encode frame-level speaker activity using four classes (STNO): $\mathcal{S}$ (silence), $\mathcal{T}$ (only target speaker active), $\mathcal{N}$ (only non-target speaker active), and $\mathcal{O}$ (target speaker overlaps with other speaker).

To effectively condition the model on these masks, DiCoW introduces the frame-level diarization-dependent transformation (FDDT) layer~\cite{polok2024dicowdiarizationconditionedwhispertarget}. Let $\mathbf{h}^l \in \mathbb{R}^{d_m \times T}$ denote the input to Encoder's $l$-th Transformer layer. FDDT applies four class-specific affine transformations, weighted by the corresponding STNO probabilities:

\begin{align}
\label{eq:FDDT}
\hat{\mathbf{h}}^l_t = &\left( \mathbf{W}_{\mathcal{S}}^l \mathbf{h}^l_t + \mathbf{b}_{\mathcal{S}}^l \right) p^t_{\mathcal{S}} + 
\left( \mathbf{W}_{\mathcal{T}}^l \mathbf{h}^l_t + \mathbf{b}_{\mathcal{T}}^l \right) p^t_{\mathcal{T}}  \nonumber \\
 +&\left( \mathbf{W}_{\mathcal{N}}^l \mathbf{h}^l_t + \mathbf{b}_{\mathcal{N}}^l\right) p^t_{\mathcal{N}} + 
\left( \mathbf{W}_{\mathcal{O}}^l \mathbf{h}^l_t + \mathbf{b}_{\mathcal{O}}^l \right) p^t_{\mathcal{O}},
\end{align}
where $p^t_{\mathcal{S}}$, $p^t_{\mathcal{T}}$, $p^t_{\mathcal{N}}$ and $p^t_{\mathcal{O}}$ correspond to STNO probabilities derived from standalone diarization at time $t$. This convex combination allows the model to adjust its internal representations depending on the speaker context. See~\cite{polok2024dicowdiarizationconditionedwhispertarget} for additional details. In this work, we used only the oracle diarization derived from human annotations, i.e. the transformation simplifies to selecting the corresponding class-specific transformation.
While these oracle masks result in one-hot vectors, the FDDT mechanism natively supports soft assignments from automatic diarization systems without architectural changes.

\vspace{-1em}
\subsection{Multi-talker ASR with speaker-attributed Whisper}
\vspace{-0.5em}
\label{sec:mt_sot}

To adapt Whisper for speaker-attributed (SA) ASR, we fine-tune the model using Serialized Output Training (SOT)~\cite{kanda20b_interspeech,Kanda2021MultiTalkerASR}, which enables a single-decoder AED model to produce interleaved transcriptions of multiple speakers.
Unlike in previous SOT-based Whisper work~\cite{li23o_interspeech}, we explicitly model speaker identities and timing by extending Whisper’s standard tokens $\mathcal{V}$ with speaker-aware timestamp tokens $\mathcal{W} \times \mathcal{U}$, e.g. $\langle|\text{s1\_2.2}|\rangle$ denote segment related to speaker 1 starting/ending at 2.2 seconds (relative to 30\,s segment).
We refer to those special tokens as \textit{speaker-timestamps}.
This resembles the approach of SLIDAR~\cite{cornell2024one}, with the key difference that timestamps and speaker labels are jointly encoded as unified tokens.

The order of speaker-attributed segments follows their onset time (FIFO). Speaker labels are assigned consistently throughout the recording based on external diarization, i.e. transcriptions for speaker $u$ correspond always to the same speaker in the entire recording.
Within each speaker's turn, timestamps are constrained to progress monotonically, while transitions to another speaker allow rolling back in time, effectively modeling overlapping speech.

\begin{figure}[htb!]
    \centering
    \includegraphics[width=0.7\columnwidth]{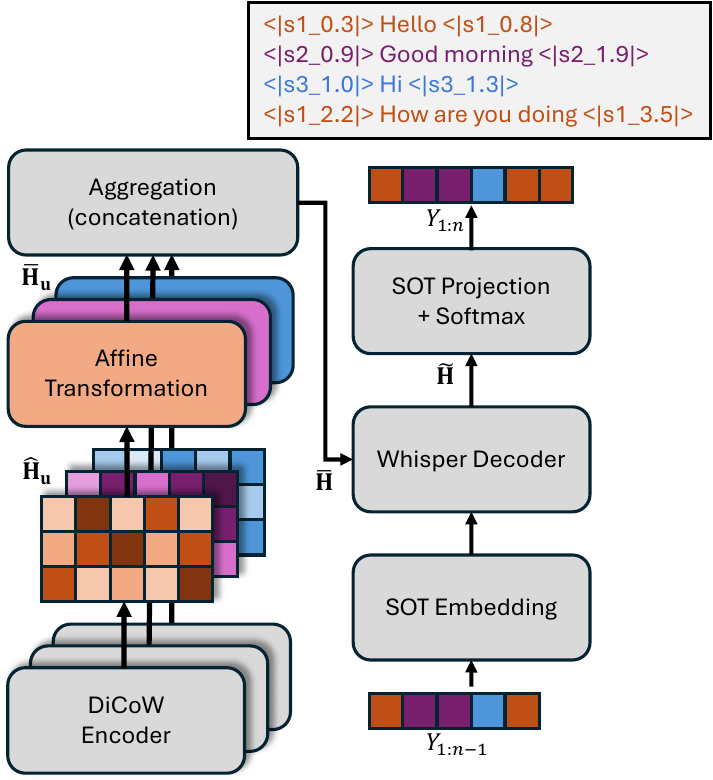}
    \label{fig:dicow_encoder}
    \caption{Overall architecture of proposed SA-DiCoW.}
    \label{fig:sa_dicow}
\end{figure}

\vspace{-1em}
\subsection{Speaker-attributed Diarization-Conditioned Whisper}
\vspace{-0.5em}
\label{sec:sa_dicow}
This section introduces our proposed Speaker-Attributed Diarization-Conditioned Whisper (SA-DiCoW) model, as shown in Figure~\ref{fig:sa_dicow}.
Given an input mixture $\mathbf{X}$ and a diarization-derived STNO mask $\mathbf{M}_u$ for each speaker $u$, we obtain speaker-specific encoder representations by running the DiCoW encoder separately for each speaker:
\begin{equation}
    \mathbf{\hat{H}}_u = \text{DiCoW Encoder}(\mathbf{X}, \mathbf{M}_u),\quad \mathbf{\hat{H}}_u \in \mathbb{R}^{d_m \times T}.
\end{equation}
These outputs, which we refer to as \textit{speaker-channel} embeddings, provide time-aligned, speaker-conditioned representations of the mixture.
To incorporate speaker identity into these embeddings, we apply a learned affine transformation to each speaker-channel embedding:
\begin{equation}
    \mathbf{\Bar{H}}_u = \mathbf{W}_u \mathbf{\hat{H}}_u + \mathbf{b}_u,\quad \mathbf{W}_u \in \mathbb{R}^{d_m \times d_m}, \mathbf{b}_u \in \mathbb{R}^{d_m}.
    \label{eq:speaker_transform}
\end{equation}
This transformation effectively injects global speaker information into the embeddings, which should help the model assign correct speaker labels throughout the recording.

Next, we stack (concatenate) all speaker-channel embeddings into a unified encoder representation:
\begin{equation}
    \label{eq:aggregation}
    \mathbf{\Bar{H}} = \cat{u=1}{|\mathcal{U}|} \mathbf{\Bar{H}}_u,\quad \mathbf{\Bar{H}} \in \mathbb{R}^{d_m \times T\cdot|\mathcal{U}|}
\end{equation}
We explored also other aggregation strategies, including weighted sum, average, masked average (based on the diarization mask). Empirically, time-wise concatenation performs the best, since it preserves original representations across speakers.


During decoding, each standard input token (i.e. $y_n \in \mathcal{V}$) is embedded as usual. 
If the token represents a speaker-timestamp (i.e. $y_n \in \mathcal{W} \times \mathcal{U}$), it is first embedded as the standard Whisper timestamp, then passed through a speaker-specific affine transformation to encode the speaker identity implicitly in the decoder (cf. \eqref{eq:speaker_transform}).
The Whisper decoder, largely unchanged, then processes the modified token embeddings along with encoder embeddings $\mathbf{\Bar{H}}$ to produce decoder hidden states $\tilde{\mathbf{H}} \in \mathbb{R}^{d_m \times N}$, which are projected into three output distributions:
\begin{equation}
\mathbf{o}_n^{(t)} \;=\; \mathbf{W}^{(t)} \, \tilde{\mathbf{H}}_n + \mathbf{b}^{(t)},
\qquad
t \in \{\text{lex}, \text{time}, \text{spk}\},
\label{eq:spk_time_logits_compact}
\end{equation}
where $\mathbf{o}^{\text{lex}}_n$ are the logits over the standard tokens, 
$\mathbf{o}^{\text{spk}}_n$ are the logits over speaker identities, 
$\mathbf{o}^{\text{time}}_n$ are the logits over timestamps, and 
$\mathbf{o}^{\text{spk-time}}_{n,uw} = \mathbf{o}^{\text{spk}}_{n,u} + \mathbf{o}^{\text{time}}_{n,w}$ are the combined logits used for speaker-timestamp tokens for speaker $u$ and time-stamps $w$ at time step $n$.
This formulation allows the model to generate speaker-attributed transcripts with accurate timestamps while maintaining compatibility with Whisper's original decoding mechanisms, aside from minimal extensions necessary to encode speaker identities.

\label{sec:sa_dicow_params}
Overall, this architecture effectively leverages diarization to structure the encoder input and guide the decoding process, enabling robust speaker-attributed transcription with only minimal modifications to the original Whisper architecture. The newly introduced model parameters are initialized to identity mappings, ensuring that the model behaves like Whisper at the beginning of training. 
\section{Experimental Setup}
\vspace{-0.25em}
All experiments are conducted using single-channel audio and oracle diarization derived from the available annotations.
\vspace{-1em}
\subsection{Datasets}
\vspace{-0.5em}
We train our proposed model using only English multi-talker datasets: NOTSOFAR~\cite{vinnikov24_interspeech}, AMI~\cite{ami_corpus}, and LibriMix~\cite{Cosentino2020LibriMixAO}. Data preparation is handled using Lhotse recipes~\cite{zelasko2021lhotse}, with minor modifications to ensure that all segments comply with the 30-second input constraint of Whisper.

\noindent\textbf{NOTSOFAR} is a recently released Microsoft multi-speaker meeting data set. 
Each meeting averages 6 minutes in duration, involving 4-8 participants and totaling 35 unique speakers. 
The recordings reflect a wide range of real-world acoustic conditions and conversational styles.
The audio was captured using a proprietary device that integrates speech enhancement techniques such as beamforming, dereverberation, and denoising.
Although the dataset comes also with simulated training data, we only use real recordings.

\noindent\textbf{AMI} is a well-established benchmark for meeting transcription. We use audio from the single distant microphone (SDM) and also individual headset mix (IHM). Only the first channel of the microphone array is used to maintain consistency with our single-channel setup.

\noindent\textbf{LibriMix} is a synthetic dataset derived from LibriSpeech~\cite{librispeech}, containing artificially mixed speech from two (Libri2Mix) or three (Libri3Mix) speakers in a left-aligned manner. Thus, the shorter source speech entirely overlaps with the longer one from the start.

\vspace{-1em}
\subsection{Evaluation and metrics}
\vspace{-.5em}
We report performance using cpWER (Concatenated minimum-Permutation Word Error Rate), as implemented in the meeteval toolkit~\cite{Neumann2023MeetEval}. The cpWER metric extends standard WER by accounting for both word recognition and diarization errors, making it particularly suitable for speaker-attributed ASR evaluation.
This metric has been widely adopted in the multi-talker ASR community, including in recent CHiME evaluations~\cite{cornell2024chime}.
The evaluation is conducted using long-form decoding with beam size of 10 via Hugging Face’s Transformers library~\cite{wolf-etal-2020-transformers}. Although the model is trained on 30-second segments, at inference we process continuous long-form audio by decoding it in consecutive 30-second chunks.
For a fair comparison with the DiCoW model, we do not apply CTC rescoring, as our proposed SA-DiCoW model does not use it either.

\vspace{-1em}
\subsection{Model settings and training}
\vspace{-.5em}
Our proposed SA-DiCoW can model up to 8 speakers, i.e. the vocabulary contains $|\mathcal{V}|=50\,364$ standard Whisper tokens and $|\mathcal{U}| \times |\mathcal{W}|=8\times 1501$ speaker-timestamp tokens.
The proposed model is initialized from a publicly available Diarization-Conditioned Whisper (DiCoW) checkpoint~\cite{polok2024targetspeakerasrwhisper}, which is based on Whisper-large-v3-turbo~\cite{radford2023whisper}. Overall, the model comprises approximately 918M trainable parameters.
Our codebase, together with the configurations of each experiment, is available on our GitHub\footnote{
\url{https://github.com/BUTSpeechFIT/SOT-DiCoW.git}
}.

Training is conducted in two stages.
In the first stage, all original Whisper encoder and decoder parameters are frozen, and only the newly introduced components (cf. Section~\ref{sec:sa_dicow_params}) are trained for 1\,000 steps using the AdamW optimizer with a learning rate of 2e-4 and a linear warm-up over the first 500 steps.
In the second stage, the full model is fine-tuned by unfreezing the Whisper parameters and applying a reduced learning rate of 2e-6 to the pre-trained weights.
This two-phase training strategy helps retain Whisper’s original linguistic capabilities while adapting it to multi-talker scenarios.
All experiments were conducted using four AMD MI250x GPUs. Each device processed a batch size of 1, and gradient accumulation was used to achieve an effective batch size of 192 per model update. The model typically converged after approximately 5\,000 training steps.

To prevent the model from memorizing specific mappings between speaker tags and speaker identities, we propose a speaker-order augmentation strategy. During training, we randomly permute the speaker labels assigned by diarization, encouraging the model to disassociate token identities from fixed speaker roles. This forces the model to learn speaker-specific representations solely based on which encoder embeddings the model currently attend to (cf. Section~\ref{sec:sa_dicow}). 
\vspace{-1.5em}
\section{Results}
\vspace{-0.5em}
\subsection{Impact of encoder embedding aggregation}
\vspace{-0.5em}

We analyze the impact of encoder embedding aggregation strategies on cpWER. These strategies correspond to different instantiations of the aggregation function in~\eqref{eq:aggregation}, listed in Table~\ref{tab:aggregation}.
In Libri2Mix, the differences are minor, with cpWER ranging from 4.6\,\% to 4.8\,\%. Concatenation performs best at 3.9\,\%, due to its ability to preserve temporal and speaker-specific patterns even in clean, fully overlapped conditions.

\begin{table}[htb!]
    \centering
    \caption{cpWER comparison of different aggregation methods of Encoder's embeddings on LibriMix Clean with 2 speakers and NOTSOFAR with 4-8 speakers.}
    \begin{tabular}{c|c|c}
        \toprule
        \textbf{Aggregation} & \textbf{Libri2Mix} & \textbf{NOTSOFAR} \\
        \midrule
        weighted sum & 4.8 & 59.1\\
        average &  4.6 & 50.2\\
        masked average & 4.6 & 47.4 \\
        concatenation & 3.9 & 21.0\\
        \bottomrule
    \end{tabular}
    \label{tab:aggregation}
\end{table}

In NOTSOFAR, which contains realistic conversations of 4 to 8 speakers per meeting, the aggregation strategy plays a critical role.
Concatenation achieves a cpWER of 21.0\,\%, significantly outperforming the the other approaches.
This represents a relative reduction of 64\,\% in cpWER w.r.t. the baseline weighted sum.
The superior performance of concatenation is due to its ability to preserve the temporal and speaker-specific structure of the embeddings.
In contrast, averaging-based methods attenuate the acoustic information across multiple speakers, especially in meetings with higher speaker counts, leading to a loss of speaker identity cues and degraded recognition accuracy.

\vspace{-1em}
\subsection{Ablation study on improving speaker label assignment}
\vspace{-.5em}

A major source of errors in multi-talker ASR arises from incorrect speaker label assignment.
Table~\ref{tab:dicow_vs_sotdicow_errors} breaks down these errors for the original DiCoW and our SOT-based approach, showing that most cpWER arises from omission (missed speech segments) and leakage (incorrect speaker attributions) related errors (cf. \cite{raj23_surt2} for details).
While DiCoW avoids label confusion by decoding each speaker independently relying on diarization.
Our proposed SA-DiCoW, on the contrary, decodes all speakers in a single sequence, theoretically allowing better overlap handling.

\begin{table}[htb!]
    \centering
    \caption{Error decomposition on NOTSOFAR. Leakage (\#L) and omission (\#O) related errors depicts absolute speaker-attributed errors w.r.t to number of reference words in\,\%.}
    \begin{tabular}{r|c|ccc|cc}
        \toprule
                  & \textbf{cpWER} & \textbf{\#S} & \textbf{\#D} & \textbf{\#I} & \textbf{\#L}  & \textbf{\#O}\\
        \midrule
        DiCoW     &  18.0 & { }9.4  & 3.4 & 5.1 &   { }3.5  &   2.1\\
        SA-DiCoW &  21.0 & 10.0    & 5.9 & 5.1 &   11.0    &   3.1\\
        \bottomrule
    \end{tabular}
    \label{tab:dicow_vs_sotdicow_errors}
\end{table}

To address speaker assignment errors, we increase the cross-entropy loss weight 5-times on speaker timestamp tokens, encouraging the model to more reliably detect speaker changes and segment boundaries.
This approach leads to consistent improvements in cpWER, as shown in Table~\ref{tab:improved_speaker_labels}.

\begin{table}[htb!]
    \centering
    \caption{Impact of improved speaker lables on cpWER for LibriSpeech Test-Other, LibriMix Test-Clean and the NOTSOFAR.}
    \begin{tabular}{r|c|cc|c}
        \toprule
        {}      & \textbf{LS-Other} & \multicolumn{2}{c|}{\textbf{LibriMix}} & \textbf{NSF}\\
        {}      & 1spk & 2spk & 3spk & 4-8spk\\
        \midrule
        DiCoW       & 4.9           & 4.8  & 32.1   & 18.0  \\
        SA-DiCoW    & 5.1         & 3.9   & 18.0  & 21.0  \\
        { }+ spk loss & 5.0       & 3.4   & 17.2  & 20.8  \\
        \bottomrule
    \end{tabular}
    \label{tab:improved_speaker_labels}
\end{table}

For reference, DiCoW, that is, a target-speaker model that decodes each speaker independently, achieves 18.4\,\% cpWER in NOTSOFAR, outperforming our best SA-DiCoW model (20.8\,\%).
However, on LibriMix with 3 speakers, DiCoW obtains 32.1\,\% cpWER, while our model achieves 17.2\,\%.
This drop in DiCoW’s performance on LibriMix is expected: in fully overlapped conditions, the STNO masks for different speakers become ambiguous, giving the model little guidance on which speaker to transcribe. These results validate the design of SA-DiCoW, demonstrating that access to global conversational context is indeed beneficial, especially in overlap-heavy conditions.

\vspace{-1.5em}
\subsection{Analysis of cross-attention weights}
\vspace{-.5em}

\begin{figure}
    \centering
    \includegraphics[width=1.0\linewidth]{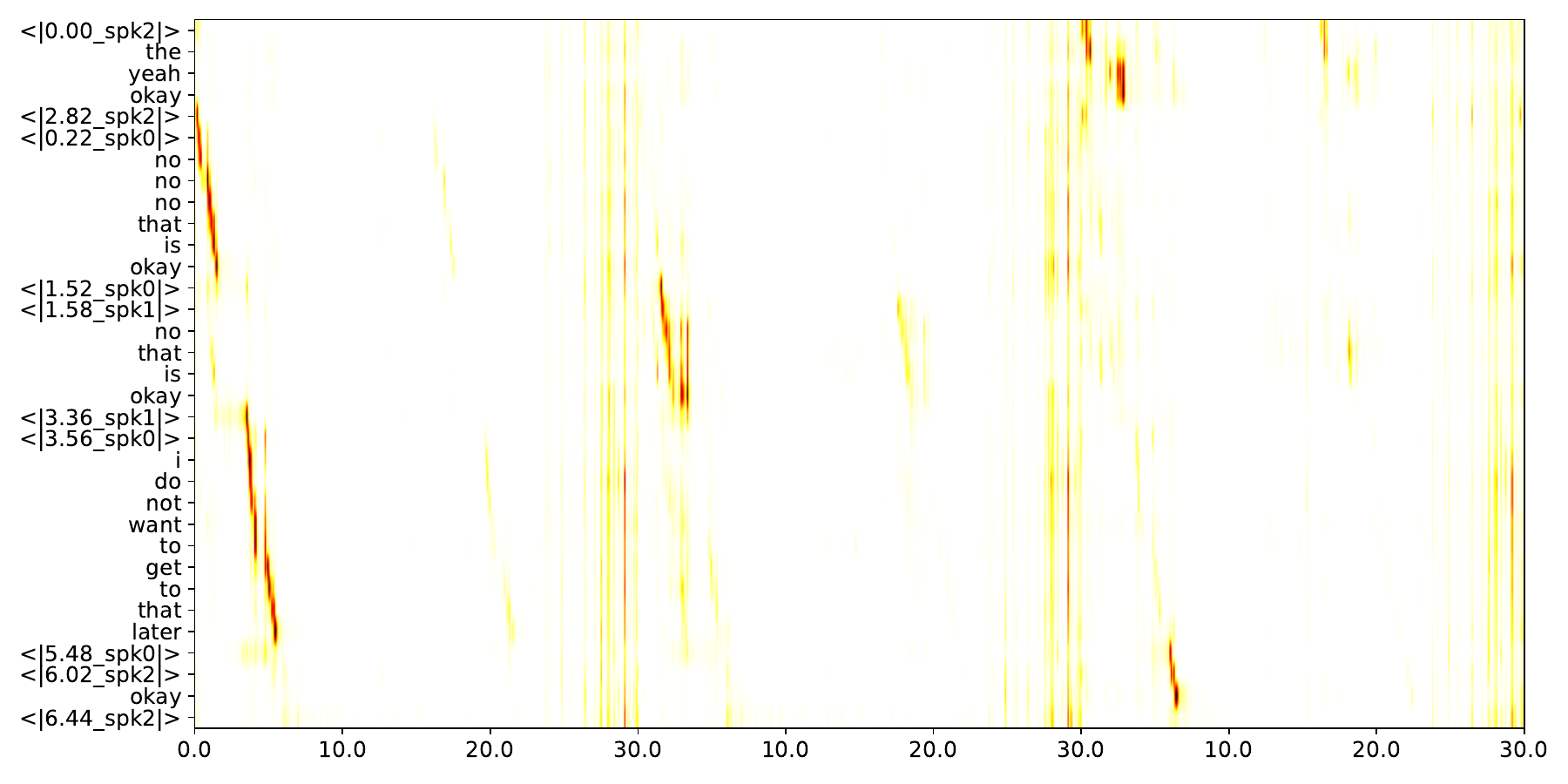}
    \vspace{-2mm}
    \caption{Example of cross-attentions from last decoder layer: x-axis depicts time in seconds, y-axis shows tokens.}
    \label{fig:x_attention1}
\end{figure}

To understand how the model uses speaker information during decoding, we visualize the cross-attention weights between the decoder and the speaker-channel embeddings from DiCoW encoder. Figure~\ref{fig:x_attention1} reveals structured and interpretable attention patterns, suggesting that the model dynamically switches between relevant speaker channels when emitting tokens. The attention visualization was generated from a random utterance sampled from the AMI corpus.

Specifically, the visualization shows the cross-attention from the last decoder layer, with all attention heads averaged to produce a single heatmap. This aggregated view highlights how the decoder’s focus shifts across different parts of the encoder sequence as it attends to different speakers. Because the encoder was constructed by concatenating speaker-channel embeddings—one per diarized segment—the attention weights jump between different speaker segments as the decoder aligns itself to the correct speaker when generating each token. This behavior underscores the model’s ability to dynamically incorporate speaker information from multiple speaker-channel embeddings, even in highly overlapped segments.

\vspace{-1em}
\subsection{Comparison with prior speaker-attributed ASR systems}
\vspace{-.5em}

Table~\ref{tab:prior_works} compares our proposed SA-DiCoW model with existing speaker-attributed ASR systems on AMI-SDM and AMI-IHM-MIX with oracle diarization.
On AMI-SDM, our method achieves (18.1\,\%) cpWER, outperforming previous works such as Cornell et al. (21.1\,\%)~\cite{cornell2024one} and Li et al. (21.2\,\%)~\cite{li23o_interspeech}.
This underscores the effectiveness of our approach in handling challenging far-field conversational data.
For reference, the original target-speaker DiCoW model achieves the lowest cpWER on both datasets.

On AMI-IHM-MIX, our model achieves 14.4\,\% cpWER, outperforming Wang et al. (26.6\,\%), and approaching the performance of SLIDAR (11.5\,\%) from~\cite{cornell2024one}. While SLIDAR performs best on IHM, it benefits from extensive training on synthetic mixtures, including augmented AMI data. In contrast, our model is fine-tuned on a combination of real conversational data (AMI-SDM and NOTSOFAR) and artificial data (Libri2Mix), which may contribute to stronger generalization to real-world scenarios, as suggested by these results. 
Overall, these findings confirm that our SA-DiCoW is competitive with the current state-of-the-art and particularly effective on realistic conversational benchmarks.


\begin{table}[htb!]
    \centering
    \caption{cpWER comparison with related works. 
    Note, the cpWER from Wang et al.\,\cite{wang_icassp25_metacat} marked with $\star$ was achieved on 10-20s long segments, i.e. the model is not penalized for speaker confusion across segments.}
    \begin{tabular}{r|c|c}
        \toprule
        {}                          & \textbf{AMI-SDM} & \textbf{AMI-IHM-MIX} \\\midrule 
        Cornell et al.\,\cite{cornell2024one}    & 21.1    & 11.5   \\ 
        Wang et al.\,\cite{wang_icassp25_metacat}&  -      & 22.8$^\star$   \\ 
        Li et al.\,\cite{li23o_interspeech}      & 21.2    &  -     \\ 
        DiCoW                                    & 16.3    &  13.1      \\\midrule 
        SA-DiCoW                         & 18.1    & 14.4   \\ 
        \bottomrule
    \end{tabular}
    \label{tab:prior_works}
\end{table}
\vspace{-1.5em}
\section{Conclusion}
\vspace{-.5em}
In this paper, we present a modified Whisper-based architecture for multi-talker speech recognition that integrates target-speaker modeling with serialized output training. Using a DiCoW encoder to extract target-speaker embeddings and concatenating them into a unified representation, our SA-DiCoW enables joint decoding of overlapping speech streams. This design allows the decoder to condition on the context of all speakers simultaneously, producing serialized transcriptions enriched with speaker tags and timestamps.

Our experimental results demonstrated that the proposed model outperforms existing SOT-based approaches on synthetic mixtures (e.g., LibriMix), achieving lower cpWER by effectively modeling speaker transitions in highly overlapped conditions. However, on real-world conversational datasets like AMI and NOTSOFAR, DiCoW established target-speaker ASR system yields superior performance, indicating that separate decoding per speaker remains advantageous in highly challenging meeting scenarios. This highlights the ongoing trade-offs between joint and separate speaker modeling in multi-talker ASR, depending on the level of overlap, background noise, and conversational complexity.


Future work will focus on improving speaker label assignment, particularly in challenging meeting datasets, by exploring speaker-aware training objectives. We also plan to investigate the potential of semi-supervised learning to reduce the reliance on oracle diarization, as well as explore adaptive speaker embedding mechanisms that can dynamically adjust to different conversational contexts. Overall, our work highlights the promise of combining target-speaker modeling and serialized output training to advance the state of MT-ASR. 

\section{Acknowledgements}
The work was supported by Czech Ministry of Culture NAKI III project JARIN (DH23P03OVV010) and by Ministry of Education, Youth and Sports of the Czech Republic (MoE) through the OP JAK project "Linguistics, Artificial Intelligence and Language and Speech Technologies: from Research to Applications" (ID:CZ.02.01.01/00/23\_020/0008518). Computing on IT4I supercomputer was supported by MoE through the e-INFRA CZ (ID:90254).


\bibliographystyle{IEEEbib}
\bibliography{references}

@article{polok2024dicowdiarizationconditionedwhispertarget,
title = {DiCoW: Diarization-conditioned Whisper for target speaker automatic speech recognition},
journal = {{Computer Speech \& Language}},
volume = {95},
pages = {101841},
year = {2026},
issn = {0885-2308},
doi = {https://doi.org/10.1016/j.csl.2025.101841},
author = {Alexander Polok and Dominik Klement and Martin Kocour and Jiangyu Han and Federico Landini and Bolaji Yusuf and Matthew Wiesner and Sanjeev Khudanpur and Jan Černocký and Lukáš Burget},
}

@article{chen2022wavlm,
  title={{WavLM}: Large-Scale Self-Supervised Pre-Training for Full Stack Speech Processing},
  author={Chen, Sanyuan and Wang, Chengyi and Chen, Zhengyang and Wu, Yu and Liu, Shujie and Chen, Zhuo and Li, Jinyu and Kanda, Naoyuki and Yoshioka, Takuya and Xiao, Xiong and others},
  journal={IEEE Journal of Selected Topics in Signal Processing},
  volume={16},
  number={6},
  pages={1505--1518},
  year={2022},
  publisher={IEEE}
}

@inproceedings{cornell2024one,
  title={One Model to Rule Them All? {Towards} End-to-End Joint Speaker Diarization and Speech Recognition},
  author={Cornell, Samuele and Jung, Jee-weon and Watanabe, Shinji and Squartini, Stefano},
  booktitle={2024 ICASSP},
  pages={11856--11860},
  year={2024},
  organization={IEEE}
}

@inproceedings{cornell2024chime,
  title={The {CHiME-8} {DASR} Challenge for Generalizable and Array Agnostic Distant Automatic Speech Recognition and Diarization},
  author    = {Samuele Cornell and Tae Jin Park and He Huang and Christoph Boeddeker and Xuankai Chang and Matthew Maciejewski and Matthew S Wiesner and Paola Garcia and Shinji Watanabe},
  year      = {2024},
  booktitle = {8th International Workshop CHiME 2024},
  pages     = {1--6},
  doi       = {10.21437/CHiME.2024-1},
}

@article{Cosentino2020LibriMixAO,
  title={{LibriMix}: An Open-Source Dataset for Generalizable Speech Separation},
  author={Joris Cosentino and Manuel Pariente and Samuele Cornell and Antoine Deleforge and Emmanuel Vincent},
  journal={arXiv: Audio and Speech Processing},
  year={2020},
  url={https://api.semanticscholar.org/CorpusID:218862876}
}

@inproceedings{ami_corpus,
author = {Carletta, Jean and Ashby, Simone and Bourban, Sebastien and Flynn, Mike and Guillemot, Mael and Hain, Thomas and Kadlec, Jaroslav and Karaiskos, Vasilis and Kraaij, Wessel and Kronenthal, Melissa and Lathoud, Guillaume and Lincoln, Mike and Lisowska Masson, Agnes and Mccowan, Iain and Post, Wilfried and Reidsma, Dennis and Wellner, Pierre},
year = {2005},
month = {07},
pages = {},
title = {The {AMI} meeting corpus: A pre-announcement},
isbn = {978-3-540-32549-9},
journal = {Lecture Notes in Computer Science},
doi = {10.1007/11677482_3}
}

@inproceedings{hsu2021hubert,
  title={{HuBERT}: How much can a bad teacher benefit {ASR} pre-training?},
  author={Hsu, Wei-Ning and Tsai, Yao-Hung Hubert and Bolte, Benjamin and Salakhutdinov, Ruslan and Mohamed, Abdelrahman},
  booktitle={2021 ICASSP},
  pages={6533--6537},
  year={2021},
  organization={IEEE}
}

@inproceedings{kanda20b_interspeech,
  title     = {Serialized Output Training for End-to-End Overlapped Speech Recognition},
  author    = {Naoyuki Kanda and Yashesh Gaur and Xiaofei Wang and Zhong Meng and Takuya Yoshioka},
  year      = {2020},
  booktitle = {Interspeech 2020},
  pages     = {2797--2801},
  doi       = {10.21437/Interspeech.2020-999},
  issn      = {2958-1796},
}

@inproceedings{Kanda2021MultiTalkerASR,
  title     = {Large-Scale Pre-Training of End-to-End Multi-Talker {ASR} for Meeting Transcription with Single Distant Microphone},
  author    = {Naoyuki Kanda and Guoli Ye and Yu Wu and Yashesh Gaur and Xiaofei Wang and Zhong Meng and Zhuo Chen and Takuya Yoshioka},
  year      = {2021},
  booktitle = {Interspeech 2021},
  pages     = {3430--3434},
  doi       = {10.21437/Interspeech.2021-102},
  issn      = {2958-1796},
}

@inproceedings{kanda21b_interspeech,
  title     = {End-to-End Speaker-Attributed {ASR} with Transformer},
  author    = {Naoyuki Kanda and Guoli Ye and Yashesh Gaur and Xiaofei Wang and Zhong Meng and Zhuo Chen and Takuya Yoshioka},
  year      = {2021},
  booktitle = {Interspeech 2021},
  pages     = {4413--4417},
  doi       = {10.21437/Interspeech.2021-101},
  issn      = {2958-1796},
}

@INPROCEEDINGS{wang_icassp25_metacat,
  author={Wang, Jinhan and Wang, Weiqing and Dhawan, Kunal and Park, Taejin and Kim, Myungjong and Medennikov, Ivan and Huang, He and Koluguri, Nithin and Balam, Jagadeesh and Ginsburg, Boris},
  booktitle={ICASSP 2025 - 2025 ICASSP}, 
  title={{META-CAT}: Speaker-Informed Speech Embeddings via Meta Information Concatenation for Multi-talker {ASR}}, 
  year={2025},
  volume={},
  number={},
  pages={1-5},
  doi={10.1109/ICASSP49660.2025.10889841}}

@inproceedings{li23o_interspeech,
  title     = {Adapting Multi-Lingual {ASR} Models for Handling Multiple Talkers},
  author    = {Chenda Li and Yao Qian and Zhuo Chen and Naoyuki Kanda and Dongmei Wang and Takuya Yoshioka and Yanmin Qian and Michael Zeng},
  year      = {2023},
  booktitle = {Interspeech 2023},
  pages     = {1314--1318},
  doi       = {10.21437/Interspeech.2023-1276},
  issn      = {2958-1796},
}

@inproceedings{meng24c_interspeech,
  title     = {Empowering {Whisper} as a Joint Multi-Talker and Target-Talker Speech Recognition System},
  author    = {Lingwei Meng and Jiawen Kang and Yuejiao Wang and Zengrui Jin and Xixin Wu and Xunying Liu and Helen Meng},
  year      = {2024},
  booktitle = {Interspeech 2024},
  pages     = {4653--4657},
  doi       = {10.21437/Interspeech.2024-971},
  issn      = {2958-1796},
}

@inproceedings{Neumann2023MeetEval,
  author    = {T. v. Neumann and C. B. Boeddeker and M. Delcroix and R. Haeb-Umbach},
  title     = {{MeetEval}: A Toolkit for Computation of Word Error Rates for Meeting Transcription Systems},
  booktitle = {Proceedings of the 7th International Workshop CHiME 2023},
  pages     = {27--32},
  year      = {2023},
  doi       = {10.21437/CHiME.2023-6},
}

@INPROCEEDINGS{polok2024targetspeakerasrwhisper,
  author={Polok, Alexander and Klement, Dominik and Wiesner, Matthew and Khudanpur, Sanjeev and Černocký, Jan and Burget, Lukáš},
  booktitle={2025 ICASSP},
  title={Target Speaker {ASR} with {Whisper}},
  year={2025},
  volume={},
  number={},
  pages={1-5},
  doi={10.1109/ICASSP49660.2025.10887683}}

@inproceedings{radford2023whisper,
  title={Robust speech recognition via large-scale weak supervision},
  author={Radford, Alec and Kim, Jong Wook and Xu, Tao and Brockman, Greg and McLeavey, Christine and Sutskever, Ilya},
  booktitle={International conference on machine learning},
  pages={28492--28518},
  year={2023},
  organization={PMLR}
}

@article{raj23_surt2,
author = {Raj, Desh and Povey, Daniel and Khudanpur, Sanjeev},
title = {{SURT 2.0}: Advances in Transducer-Based Multi-Talker Speech Recognition},
year = {2023},
issue_date = {2023},
publisher = {IEEE Press},
volume = {31},
issn = {2329-9290},
url = {https://doi.org/10.1109/TASLP.2023.3318398},
doi = {10.1109/TASLP.2023.3318398},
journal = {IEEE/ACM Transactions on Audio, Speech, and Language Processing},
month = {sep},
pages = {3800–3813},
numpages = {14}
}

@inproceedings{vaswani2017attention,
 author = {Vaswani, Ashish and Shazeer, Noam and Parmar, Niki and Uszkoreit, Jakob and Jones, Llion and Gomez, Aidan N and Kaiser, \L ukasz and Polosukhin, Illia},
 booktitle = {Advances in Neural Information Processing Systems},
 editor = {I. Guyon and U. Von Luxburg and S. Bengio and H. Wallach and R. Fergus and S. Vishwanathan and R. Garnett},
 pages = {},
 publisher = {Curran Associates, Inc.},
 title = {Attention is All you Need},
 volume = {30},
 year = {2017}
}

@inproceedings{vinnikov24_interspeech,
  title     = {{NOTSOFAR-1} Challenge: New Datasets, Baseline, and Tasks for Distant Meeting Transcription},
  author    = {Alon Vinnikov and Amir Ivry and Aviv Hurvitz and Igor Abramovski and Sharon Koubi and Ilya Gurvich and Shai Peer and Xiong Xiao and Benjamin Martinez Elizalde and Naoyuki Kanda and Xiaofei Wang and Shalev Shaer and Stav Yagev and Yossi Asher and Sunit Sivasankaran and Yifan Gong and Min Tang and Huaming Wang and Eyal Krupka},
  year      = {2024},
  booktitle = {Interspeech 2024},
  pages     = {5003--5007},
  doi       = {10.21437/Interspeech.2024-1788},
  issn      = {2958-1796},
}

@inproceedings{wolf-etal-2020-transformers,
    title = "Transformers: State-of-the-Art Natural Language Processing",
    author = "Thomas Wolf and Lysandre Debut and Victor Sanh and Julien Chaumond and Clement Delangue and Anthony Moi and Pierric Cistac and Tim Rault and Rémi Louf and Morgan Funtowicz and Joe Davison and Sam Shleifer and Patrick von Platen and Clara Ma and Yacine Jernite and Julien Plu and Canwen Xu and Teven Le Scao and Sylvain Gugger and Mariama Drame and Quentin Lhoest and Alexander M. Rush",
    booktitle = "In the 2020 Conference on Empirical Methods in {NLP}: System Demonstrations",
    month = oct,
    year = "2020",
    address = "Online",
    publisher = "Association for Computational Linguistics",
    pages = "38--45"
}

@inproceedings{yu2017permutation,
  title={Permutation invariant training of deep models for speaker-independent multi-talker speech separation},
  author={Yu, Dong and Kolb{\ae}k, Morten and Tan, Zheng-Hua and Jensen, Jesper},
  booktitle={2017 ICASSP},
  pages={241--245},
  year={2017},
  organization={IEEE}
}

@INPROCEEDINGS{librispeech,
  author={Panayotov, Vassil and Chen, Guoguo and Povey, Daniel and Khudanpur, Sanjeev},
  booktitle={2015 ICASSP},
  title={Librispeech: An {ASR} corpus based on public domain audio books},
  year={2015},
  volume={},
  number={},
  pages={5206-5210},
  doi={10.1109/ICASSP.2015.7178964}
}

@misc{abramovski2025summarynotsofar1challengehighlights,
      title={Summary of the {NOTSOFAR-1 Challenge}: Highlights and Learnings},
      author={Igor Abramovski and Alon Vinnikov and Shalev Shaer and Naoyuki Kanda and Xiaofei Wang and Amir Ivry and Eyal Krupka},
      year={2025},
      eprint={2501.17304},
      archivePrefix={arXiv},
      primaryClass={cs.SD},
      url={https://arxiv.org/abs/2501.17304},
}

@article{zelasko2021lhotse,
  title={Lhotse: a speech data representation library for the modern deep learning ecosystem},
  author={Żelasko, Piotr and Povey, Daniel and Trmal, Jan and Khudanpur, Sanjeev},
  journal={Proceedings of NeurIPS Workshop on Data-Centric AI},
  year={2021}
}

@inproceedings{chorowski2015_aed,
 author = {Chorowski, Jan K and Bahdanau, Dzmitry and Serdyuk, Dmitriy and Cho, Kyunghyun and Bengio, Yoshua},
 booktitle = {Advances in Neural Information Processing Systems},
 editor = {C. Cortes and N. Lawrence and D. Lee and M. Sugiyama and R. Garnett},
 pages = {},
 publisher = {Curran Associates, Inc.},
 title = {Attention-Based Models for Speech Recognition},
 volume = {28},
 year = {2015}
}

@INPROCEEDINGS{chan2015_las,
  author={Chan, William and Jaitly, Navdeep and Le, Quoc and Vinyals, Oriol},
  booktitle={2016 ICASSP}, 
  title={Listen, attend and spell: A neural network for large vocabulary conversational speech recognition}, 
  year={2016},
  volume={},
  number={},
  pages={4960-4964},
  doi={10.1109/ICASSP.2016.7472621}}

@inproceedings{meng23b_interspeech,
  title     = {Unified Modeling of Multi-Talker Overlapped Speech Recognition and Diarization with a Sidecar Separator},
  author    = {Lingwei Meng and Jiawen Kang and Mingyu Cui and Haibin Wu and Xixin Wu and Helen Meng},
  year      = {2023},
  booktitle = {Interspeech 2023},
  pages     = {3467--3471},
  doi       = {10.21437/Interspeech.2023-1422},
  issn      = {2958-1796},
}

@INPROCEEDINGS{han25_icassp,
  author={Han, Jiangyu and Landini, Federico and Rohdin, Johan and Silnova, Anna and Diez, Mireia and Burget, Lukáš},
  booktitle={ICASSP 2025}, 
  title={Leveraging Self-Supervised Learning for Speaker Diarization}, 
  year={2025},
  volume={},
  number={},
  pages={1-5},
  doi={10.1109/ICASSP49660.2025.10889475}}

@inproceedings{zmolikova20_chime,
  author="Kateřina {Žmolíková} and Martin {Kocour} and Federico Nicolás {Landini} and Karel {Beneš} and Martin {Karafiát} and Hari Krishna {Vydana} and Alicia {Lozano Díez} and Oldřich {Plchot} and Murali Karthick {Baskar} and Ján {Švec} and Ladislav {Mošner} and Vladimír {Malenovský} and Lukáš {Burget} and Bolaji {Yusuf} and Ondřej {Novotný} and František {Grézl} and Igor {Szőke} and Jan {Černocký}",
  title="BUT System for CHiME-6 Challenge",
  booktitle="Proceedings of CHiME 2020 Virtual Workshop",
  year="2020",
  pages="1--3",
  publisher="University of Sheffield",
  address="Barcelona",
  doi="10.21437/CHiME.2020-13",
  url="https://www.isca-speech.org/archive/CHiME_2020/pdfs/CHiME_2020_paper_zmolikova.pdf"
}

@ARTICLE{lu21_icassp,
  author={Lu, Liang and Kanda, Naoyuki and Li, Jinyu and Gong, Yifan},
  journal={IEEE Signal Processing Letters}, 
  title={Streaming End-to-End Multi-Talker Speech Recognition}, 
  year={2021},
  volume={28},
  number={},
  pages={803-807},
  doi={10.1109/LSP.2021.3070817}}

@misc{polok2026sedicowselfenrolleddiarizationconditionedwhisper,
      title={{SE-DiCoW}: Self-Enrolled Diarization-Conditioned Whisper}, 
      author={Alexander Polok and Dominik Klement and Samuele Cornell and Matthew Wiesner and Jan Černocký and Sanjeev Khudanpur and Lukáš Burget},
      year={2026},
      eprint={2601.19194},
      archivePrefix={arXiv},
      primaryClass={eess.AS},
      url={https://arxiv.org/abs/2601.19194}, 
}

@inproceedings{zheng-etal-2025-dncasr,
    title = "{DNCASR}: End-to-End Training for Speaker-Attributed {ASR}",
    author = "Zheng, Xianrui  and
      Zhang, Chao  and
      Woodland, Phil",
    editor = "Che, Wanxiang  and
      Nabende, Joyce  and
      Shutova, Ekaterina  and
      Pilehvar, Mohammad Taher",
    booktitle = "In the 63rd Annual Meeting of the Association for Computational Linguistics",
    month = jul,
    year = "2025",
    address = "Vienna, Austria",
    publisher = "Association for Computational Linguistics",
    url = "https://aclanthology.org/2025.acl-long.899/",
    doi = "10.18653/v1/2025.acl-long.899",
    pages = "18369--18383",
    ISBN = "979-8-89176-251-0"
}

@inproceedings{
park2025sortformer,
title={Sortformer: A Novel Approach for Permutation-Resolved Speaker Supervision in Speech-to-Text Systems},
author={Taejin Park and Ivan Medennikov and Kunal Dhawan and Weiqing Wang and He Huang and Nithin Rao Koluguri and Krishna C Puvvada and Jagadeesh Balam and Boris Ginsburg},
booktitle={Forty-second International Conference on Machine Learning},
year={2025},
url={https://openreview.net/forum?id=AyYjRvrbDx}
}
\end{document}